\documentclass[letter, twocolumn]{jpsj3}
\usepackage{txfonts}

\title{
Magnetic-Field-Independent Ultrasonic Dispersions in the Magnetically Robust Heavy Fermion System SmOs$_4$Sb$_{12}$
}

\author{
Tatsuya YANAGISAWA$^{1, 2}$, Hitoshi SAITO$^2$, Hiroyuki HIDAKA$^2$, Hiroshi AMITSUKA$^2$, Koji ARAKI$^3$,
Mitsuhiro AKATSU$^3$, Yuichi NEMOTO$^3$, Terutaka GOTO$^3$, Pei-Chun HO$^4$, Ryan E. BAUMBACH$^5$, and M. Brian MAPLE$^5$
}

\inst{
$^1$Creative Research Initiative, Hokkaido University, Sapporo 001-0021, Japan\\
$^2$Department of Physics, Hokkaido University, Sapporo 060-0810, Japan\\
$^3$Graduate School of Science and Technology, Niigata University, Niigata 950-2181, Japan\\
$^4$Department of Physics, California State University Fresno, Fresno 93740, USA\\
$^5$Department of Physics, University of California San Diego, La Jolla 92093, USA\
}

\abst{
Elastic properties of the filled skutterudite compound SmOs$_4$Sb$_{12}$ have been investigated by ultrasonic measurements. The elastic constant $C_{11}(\omega)$ shows two ultrasonic dispersions at $\sim$15 K and $\sim$53 K for frequencies $\omega$ between 33 and 316 MHz, which follow a Debye-type formula with Arrhenius-type temperature-dependent relaxation times, and remain unchanged even with applied magnetic fields up to 10 T.
The corresponding activation energies were estimated to be $E_2$ = 105 K and $E_1$ = 409 K, respectively. The latter, $E_1$, is the highest value reported so far in the Sb-based filled skutterudites. The presence of magnetically robust ultrasonic dispersions in SmOs$_4$Sb$_{12}$ implies a possibility that an emergence of a magnetically insensitive heavy fermion state in this system is associated with a novel local charge degree of freedom which causes the ultrasonic dispersion.}

\kword{filled skutterudite, heavy fermion, rattling, elastic constant, ultrasonic dispersion}

\begin{document}
\maketitle

Intermetallic compounds with cage like structures have been studied both experimentally and theoretically for several decades \cite{Keppens, Yu}. Recently, this area of research has received intense interest, since a variety of exotic physical properties are observed in these types of compounds; {\it e.g.}, heavy-electron states of heavier rare-earth (Pr, Nd, and Sm) and La based systems,
multipole ordering, unconventional superconductivity, high thermoelectric figures of merit, and anharmonic vibrations of the guest ions \cite{Sales, Chakoumakos, Sato}. It has been suggested that these exotic phenomena may originate from the characteristic cage structure in these systems.
A well-known system of intermetallic system, with a cage like structure, that exhibits a variety of correlated electron phenomena, are the filled skutterudite \cite{Brawn, Chakoumakos, Sales, Sato}. For instance, an unconventional heavy fermion (HF) state is observed for SmOs$_4$Sb$_{12}$ \cite{Yuhasz}, where the electronic specific heat coefficient $\gamma$=820~mJ/K$^2$mol is insensitive to an applied magnetic field $H~\|~$[100] up to 8 T \cite{Sanada}. This result is in contrast to the expected suppression of $\gamma$ with magnetic field for the conventional (magnetic) Kondo effect, where the heavy fermion ground state is derived from spin degrees of freedom, as is often observed for Ce- and Yb-based compounds. In contrast, the exotic HF state in SmOs$_4$Sb$_{12}$ might be explained by a novel many-body effect, such as the multichannel Kondo effect, which is derived from multipole degrees of freedom and/or local charge fluctuations originated from off-center degrees of freedom of guest ions. \cite{Hotta, Hattori}

By means of ultrasonic measurements, a frequency dependent upturn of the elastic constant with ultrasonic attenuation peak, called an ultrasonic dispersion (UD), has been observed in the cage compounds, such as $R_3$Pd$_{20}$Ge$_6$ ($R$ = La, Ce, Pr) and $R$Os$_4$Sb$_{12}$ ($R$ = La, Pr, Nd) \cite{LPG, Nemoto, CPG}. Moreover, an unusual decreasing of the elastic constant has also been observed at very low temperatures in the La based compounds. Due to the co-presence of UD and low temperature elastic softening in non-$4f$ La based compounds, the origins of these phenomena are expected to share a common root cause, originated with a novel local charge degrees of freedom. As a candidate for the origin, a model based on an off-center ionic configurations in a multi-well potential has been proposed. Here, the UD can be understood as a thermally activated off-center motion of guest ion, which was defined as e(off-center) rattlingf in our previous papers \cite{CPG, Yanagisawa}. On the contrary, recent neutron scattering experiment on PrOs$_4$Sb$_{12}$ has revealed that the Pr-nuclei density distribution with an accuracy of 0.1$\AA$ is on-center-like without anisotropy at 8 K while it is off-center-like with strong anharmonicity at room temperature.\cite{Kaneko} Therefore, it is still controversial whether the off-center configuration of guest ion is appropriate for understanding the above elastic anomalies.

Here, we note that the word `rattling' has originally been used for the localized thermal vibration of the guest atom in an over-sized host cage with large amplitude.\cite{Brawn} In the rattling systems, a low-lying optical phonon excitation of $\hbar \omega \sim$ meV has been commonly observed in several physical quantities, {\it e.g.}, Raman scattering, inelastic x-ray or neutron scattering and Einstein temperatures in the specific heat.\cite{Ogita, Tsutsui, Iwasa, Matsuhira} On the other hand, since the ultrasonic measurement with frequencies of $\sim$ MHz can monitor relatively lower energy excitations of $\hbar \omega \sim 10^{-5}$ meV, a causal linkage between the optical phonon and UD has not been confirmed thus far and still an open question. In order to shed light on the nature of the rattling phenomenon and UD, a systematic change of the characteristic parameters obtained from the UD and other physical properties among the Sb-based filled skutterudites will be reviewed in the present paper. In addition, although ultrasonic measurements of SmOs$_4$Sb$_{12}$ have already been reported\cite{Nakanishi}, there is no mention of the ultrasonic frequency dependence. The present study is focused on investigating the UD in SmOs$_4$Sb$_{12}$ in order to examine the magnetically robust novel local charge degree of freedom, which might be a possible candidate for the origin of exotic HF state in SmOs$_4$Sb$_{12}$.

Single crystals of SmOs$_4$Sb$_{12}$ were grown using a molten flux growth method. The high quality of the single crystals used in the present study were confirmed by energy dispersive x-ray spectroscopy, powder x-ray diffraction (for single crystals picked out of same sample batch), the large residual resistivity ratio (RRR (300 K/ 2 K)) of 16.5, and an observation of acoustic de Haas-van Alphen signals in the elastic constant $C_{11}$ (not shown). A cubic shaped specimen, with a length of 0.775 mm, was used for the ultrasonic measurements. A change in the sound velocity was detected by the conventional phase comparator method with pulsed ultrasound generated by LiNbO$_3$ transducers. Ultrasonic attenuation measurements were performed by recording the in-phase-(sine) and quadruture-(cosine) signals at fixed frequencies and the phase shift \cite{Wolf}. An Oxford Instruments Heliox-TL $^3$He refrigerator with a superconducting magnet was used for measurements at $T \ge$ 300 mK and $H \le$ 10 T.
\begin{figure}[t]
\begin{center}
\includegraphics[width=0.7\linewidth]{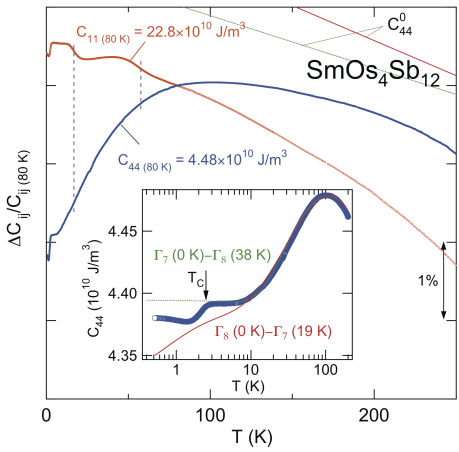}
\end{center}
\caption{Elastic consants $C_{11}$ (176 MHz) and $C_{44}$ (165 MHz) of SmOs$_4$Sb$_{12}$ as a function of temperature $T$, displayed as a relative change for comparison. Inset:  the green and red curves, displayed on a log$T$ scale, represent the fits of the quadrupole susceptibility by using the $\Gamma_7$ and $\Gamma_8$ ground states, respectively. The backgrounds of the elastic constant $C_{44}^0$ curves for the fits are displayed as the same colored-dotted lines in the main panel.}
\label{f1}
\end{figure}
\begin{figure}[t]
\begin{center}
\includegraphics[width=0.7\linewidth]{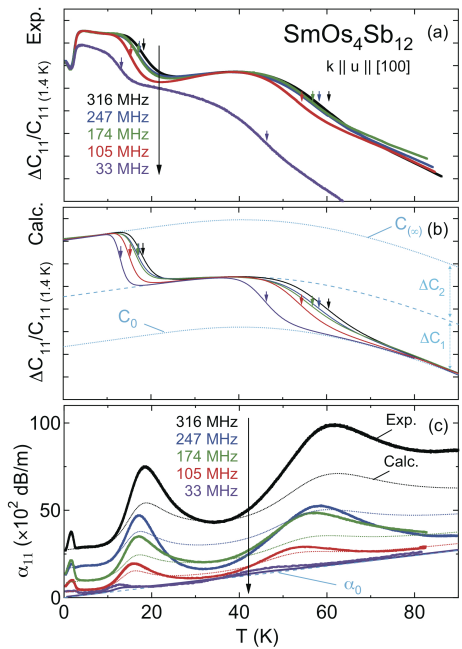}
\end{center}
\caption{(a) Relative change in the elastic constant $\Delta C_{11}/C_{11}$ as a function of temperature $T$ with frequencies varying from 33 to 316 MHz. (b) Theoretical fits to the frequency-dependent elastic constant. Arrows indicate the temperatures satisfying the resonant condition $\omega \tau \sim 1$, (c) Ultrasonic attenuation coefficient $\alpha_{11}$ as a function of temperature. Solid lines represent phenomenological calculations using the same parameters as used in the calculations in (b).}
\label{f2}
\end{figure}
%

A comparison of the elastic constants $C_{11}$ and $C_{44}$, as a function of temperature, is displayed as the relative change in $\Delta C_{\rm ij}/C_{\rm ij}$ in Fig. 1. 
Note that $C_{11} = C_B+\frac{4}{3} \frac{C_{11}-C_{12}}{2}$ consists of $\Gamma_1$-symmetry bulk modulus $C_B = \frac{C_{11}+2C_{12}}{3}$ and $\Gamma_{23}$-symmetry $\frac{C_{11}-C_{12}}{2}$ mode while $C_{44}$ mode associates with $\Gamma_4$-symmetry.
$C_{11}$ shows two upturns that are frequency-dependent and magnetic field independent (details will be described later in Figs 2 and 3). The temperatures for the appearance of these two UDs are marked by vertical dotted lines in Fig. 1. In contrast, $C_{44}$ shows neither an upturn nor a frequency dependence at these temperatures, except for the elastic softening due to the crystalline electric field (CEF) effect from $\sim100$ K to 10 K and a step-like decrease at $T_{\rm C}=2.5$ K, below which a weak ferromagnetic moment develops for SmOs$_4$Sb$_{12}$\cite{Yuhasz}.

Similar contrasting behavior between $C_{11}$ and $C_{44}$ was also observed in the $R$Os$_4$Sb$_{12}$ ($R$ = La-Nd) systems. These ultrasonic mode-dependence suggest that the UD would be derived from a symmetrical off-center charge fluctuation of the guest ion, {\it e.g.}, $\Gamma_{23}$-symmetry off-center mode for $R$Os$_4$Sb$_{12}$ and $\Gamma_5$-symmetry for $R_3$Pd$_{20}$Ge$_6$. \cite{Nemoto} We note that our results, the softening of $C_{44}$ in particular, are completely different from previous work by Nakanishi {\it et al.} \cite{Nakanishi} The major difference could simply come from the fact that higher frequency and shorter pulses, {\it i.e.}, more focused ultrasonic waves, were used in the present measurements, which help to minimize the irregular reflection of ultrasonic waves from the crack or cavity on surface or inside the crystal. A good example of this effect can be seen in a deviation of the lowest frequency (33 MHz) data of $C_{11}$ in Fig. 2 (a).

The softening of $C_{44}$ can be analyzed in terms of quadrupole susceptibility based on the CEF effect \cite{Luethi}. Thus far, two different CEF energy level schemes have been proposed for the Sm$^{3+}$ ion ($J$ = 5/2) in SmOs$_4$Sb$_{12}$ in $O_{\rm h}$ symmetry: {\it i.e.}, $\Gamma_7$(0 Ĭ)-$\Gamma_8$(38 Ĭ) and $\Gamma_8$(0 Ĭ) - $\Gamma_7$(19 Ĭ) \cite{Yuhasz, Aoki}. The inset of Fig. 1 shows theoretical fits to the experimental data below 200 K for both CEF schemes, both of which reproduce the softening of $C_{44}$ down to $\sim10$ K. The leveling off of $C_{44}$ below $\sim10$ K is described by Van Vleck-type behavior, derived from a $\Gamma_7$-doublet ground state using fitting parameters including a coupling constant for a quadrupole-strain interaction $|g| = 258$ K and for an inter-site quadrupole interaction $g'=-0.79$ K. In contrast, the scheme with a $\Gamma_8$-quartet ground state gives a fit which deviates from the $C_{44}$ data due to a Curie-type contribution from the $\Gamma_8$-quartet ground state, that persists down to $T_C$. We should, however, note that the large value of the low temperature specific heat indicates that complex many-body physics dominates the low temperature state in SmOs$_4$Sb$_{12}$.\cite{Sanada} Therefore, it is likely that 
the deviation of the $C_{44}$ data from the fit, {\it i.e.}, the shoulder for $T \lesssim$ 10 K, should probably be explained by a modified quadrupole susceptibility picture that includes a crossover from a Kondo singlet state ({\it e.g.}, as described for CePd$_2$Al$_3$)\cite{Sakai, Nemoto2} or some other exotic Kondo state ({\it e.g.}, the multichannel Kondo effect)\cite{Cox, Hattori}. Thus, the possibility of the $\Gamma_8$-quartet ground state still remains. If we skip the adjustment of the many-body effect, the fitting parameters for the $\Gamma_8$ model assume relatively larger values:  $|g|=369$ K and $g'=-1.76$ K. In the present paper, we will focus on the ultrasonic dispersions and leave the discussion of the correlated electron physics on the CEF effects for a later publication. 

Figures 2 (a) and (c) show the relative change in the elastic constant $\Delta C_{11}/C_{11}$ and the ultrasonic attenuation coefficient $\alpha_{11}$ as a function of temperature with fixed frequencies varying from 33 to 316 MHz. The two upturns in $C_{11}$, indicated by the arrows in Fig. 2(a), are found to shift to higher temperatures with increasing frequency. Maxima in $\alpha_{11}$ are also observed at $\sim53$ K and $\sim15$ K, which show the same frequency dependence as the upturns in $C_{11}$.
The frequency dependent elastic constant $C(\omega)$, $\omega$ is ultrasonic frequency, and the attenuation coefficient $\alpha(\omega)$ are well described phenomenologically, as shown in Fig. 2(b), by a Debye-type dispersion given by \cite{Yanagisawa}, $C(\omega)=C_{(\infty)}-\sum_{i=1,2} \frac{\Delta C_i}{1+\omega^2 \tau_i^2}$ and $\alpha(\omega)=\alpha_0-\sum_{i=1,2}\frac{\Delta C_i}{2\rho v_{i, \infty}^3} \frac{\omega^2 \tau_i}{1+\omega^2 \tau_i^2}$, 
where an Arrhenius-type temperature dependence of the relaxation time, $\tau_i = \tau_{0,(i)} \exp(E_i/k_BT)$, is assumed. Here, $i =$ 1 and 2 are the indices for analyzing two separated sets of UDs at $\sim 53$ K and $\sim 15$ K, respectively.
Here, $\Delta C_i$ in
%
%
the above equations
is the total change for each upturn between the high-frequency limit $C_{\infty}$ and the low-frequency limit $C_0$ of the elastic constant, which are estimated from the extrapolation of the temperature dependence of $C_{11}$ over 20 K, without the UDs and possible CEF effect (as seen below 10 K in Fig, 2(a)), as displayed as dotted curves in Fig. 2(b).
Two parameter sets of activation energy $E_i$ and attempt time $\tau_{0, (i)}$ were obtained from the phenomenological fits for the relaxation $\omega \tau \sim 1$ of Fig. 2(b) and 2(c), even though the ultrasonic attenuation peaks are often experimentally observed sharper than the phenomenological fits.~\cite{Yanagisawa} The parameters are summarized in Table 1 with the related compounds' parameters of UD and also other physical properties, including the lattice constant $a$, the Debye and Einstein temperatures ${\it \Theta}_{\rm D}$ and ${\it \Theta}_{\rm E}$, and the Sommerfeld coefficient $\gamma$.

Figure 3 shows $C_{11}$ vs $T$ at a frequency of 105 MHz and for various magnetic fields between 0 and 10 T. The UDs are unaffected by magnetic fields up to 10 T except for the field dependent CEF effect that appears in $C_{11}$ below 15 K, as can be seen in the enlarged scale in the inset of Fig. 3. Such insensitivity in the UD to magnetic fields is commonly observed in the $R$Os$_4$Sb$_{12}$ ($R$ = La-Nd) system, providing strong evidence that the UD is not magnetic but rather electric, {\it i.e.,} the novel local charge degrees of freedom, in origin.
\begin{figure}[t]
\begin{center}
\includegraphics[width=0.7\linewidth]{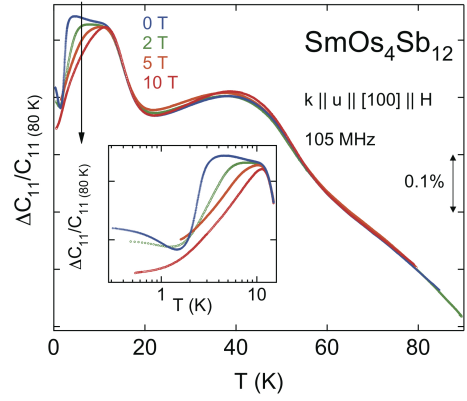}
\end{center}
\caption{Temperature dependence of the relative change of $C_{11}$ for various magnetic fields $H$ along the [001] direction. The inset displays an enlarged view below 20 K on a log$T$ scale.}
\label{f3}
\end{figure}

Figures 4 (a) and (b) show a systematic change of the UDs and an Arrhenius plot of the relaxation time $\tau_i$ vs $1/T$ for $R$Os$_4$Sb$_{12}$ ($R$ = La-Sm), respectively. The double UDs so far only have been observed in NdOs$_4$Sb$_{12}$ ($4f^3$) and SmOs$_4$Sb$_{12}$ ($4f^4$), while LaOs$_4$Sb$_{12}$ ($4f^0$) and PrOs$_4$Sb$_{12}$ ($4f^2$) show single UD. It has been pointed out that the observations of double UDs are comparable to the $^{123}$Sb-NQR measurements on $R$Os$_4$Sb$_{12}$ ($R$ = La-Sm), which show double or triple peaks in the temperature dependence of $1/T_2$ for $R$ = Pr-Sm due to some local charge fluctuation with a correlation time of $\sim10^{-6}$ sec, while LaOs$_4$Sb$_{12}$ shows only a single peak \cite{Kotegawa}.

When the rare-earth ion changes from $R$ = La to Sm, not only the guest ion's mass but also the amount of free space for the Einstein oscillator in the atomic cage increases, due to the lanthanide contraction, {\it i.e.}, the ionic radius of the guest ion decreases while the cage radius remains nearly constant for $R$Os$_4$Sb$_{12}$.\cite{Brawn} We speculate that such changes in the guest ion's conditions give rise to additional anharmonicity of the potential, {\it e.g.}, flat base or double well, for the local ionic configuration, and result the emergence of additional UD and NQR peaks in $R$ = Nd and Sm.

\begin{figure}[t]
\begin{center}
\includegraphics[width=0.9\linewidth]{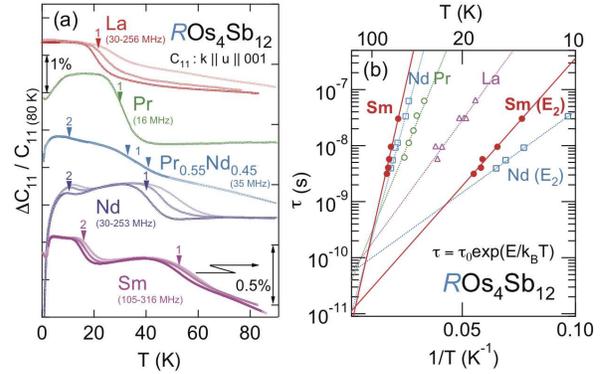}
\end{center}
\caption{(a) Comparison of the UDs that appear in elastic constant $C_{11}$ of $R$Os$_4$Sb$_{12}$ ($R$ = La-Sm) at several frequencies. Lower arrowheads with numbers 1 and 2 indicate the relaxation point $\omega \tau_i \sim 1$ for $i$ = 1 and 2, respectively. The displayed data have been shifted to eliminate overlapping with each other and the SmOs$_4$Sb$_{12}$ data are magnified three times for the $\Delta C_{11}/C_{11}$-axis. (b) Arrhenius plots of the characteristic parameters of the UDs ($\tau_{0 (i)}$, $E_i$) for $R$Os$_4$Sb$_{12}$ ($R$ = La-Sm).}
\label{f4}
\end{figure}
%
\begin{table*}[t]
\caption{\label{tab:table1}Comparison of characteristic parameters of UDs and rattling, and some physical properties of the related filled skutterudites \cite{Brawn, Nemoto, Matsuhira, Ishii}.}
\begin{tabular}{ccccccccc}
\hline
 &Lattice Constant&Debye Temp.&Einstein Temp.&Sommerfeld Coef.&Activation Energy&Attempt Time&&\\
 &$a$ (\AA)&${\it \Theta}_D$ (K)&${\it \Theta}_E$ (K)&$\gamma$ (mJ mol$^{-1}$K$^{-2}$)&$E_1$ (K)&$\tau_{0,(1)}$ (ps)&$E_2$ (K)&$\tau_{0,(2)}$ (ps)\\
\hline
LaOs$_4$Sb$_{12}$&	9.3081&	270&	60.5&	36-56&		127&	50&	-&	-\\
CeOs$_4$Sb$_{12}$&	9.3011&	304&	-&		92-180&	-&	-&	-&	-\\
PrOs$_4$Sb$_{12}$&	9.3031&	165-320&	-&		310-750&	225&	31&	-&	-\\
NdOs$_4$Sb$_{12}$&	9.2989&	255&	39&		520&	337&	7.5&	67&	51\\
SmOs$_4$Sb$_{12}$&	9.3009&	294&	40.1&	820&	409&	4.4&	105&	10\\
\hline
LaRu$_4$Sb$_{12}$&	9.2700&	275&	72.8&	47&		-&	-&	-&	-	\\
LaFe$_4$Sb$_{12}$&	9.1395&	314&	87.6&	122&	300&	31&	-&	-	\\
PrFe$_4$Sb$_{12}$&	9.1351&	-&		-&		$\sim$1000	&	360&	3&	-&	-	\\
\hline
\end{tabular}
\end{table*}
%

In Fig. 4 (b), the slope of the line is equivalent to the activation energy and the markers indicate frequencies and temperatures where the relaxation of UD meets a resonant condition $\omega \tau_i \sim 1$.
We find that the activation energy $E_1$ and Sommerfeld coefficient $\gamma$ increase and $\tau_{0, (1)}$ decreases from $R$ = La to Sm in the $R$Os$_4$Sb$_{12}$ series as summarized in Table 1. The increase $\gamma$ means that a characteristic temperature of the electronic correlation, such as Kondo temperature, will exhibit opposite (decrease) tendency in general. Therefore, the similar increase tendency of $E_1$, $E_2$ and $\gamma$, found in the present study, implies that the activation energies are not {\it directly} connected to the characteristic temperature of the electron mass-enhancement mechanism in SmOs$_4$Sb$_{12}$.

Next we compare the Einstein temperature with activation energies. If we pick out the lower activation energy of each compound for comparison, {\it i.e.} $E_1$ = 127 K for La, $E_2$ = 67 K for Nd and $E_2$ = 105 K for Sm, they seem to be qualitatively comparable to the systematic change of Einstein temperature ${\it \Theta}_E$; 60.5 K for La, 39 K for Nd and 40.1 K for Sm in Table 1. This fact implies that the optical phonon and UD could be linked with unknown pre-factor of $\sim 2$. This linkage might be due to a novel energy dissipation mechanism via conduction electron-phonon interaction, as previously proposed by Hattori and Miyake \cite{Hattori}. If there exists such conduction electron-phonon interaction, it can not be ruled out that the origin of activation energies is {\it indirectly} involved in the effective mass-enhancement mechanism, which might be mediated by the interaction between conduction electron and phonon in the present system.

Finally, we discuss a possible co-presence of Curie-type softening and UD in SmOs$_4$Sb$_{12}$. The low temperature softening as observed in LaOs$_4$Sb$_{12}$ could, however, not be found in the present study of SmOs$_4$Sb$_{12}$ because the drastic change of elastic constants due to ferromagnetic ordering at 2.5 K and the CEF effect under magnetic fields hides such features. In order to elucidate the possible ground states of the local charge degrees of freedom and extract the CEF level scheme in the present compound, further measurements, such as elastic constant ($C_{11}$-$C_{12}$)/2 and $C_{44}$ measurements in a magnetic field, are needed. In the present discussions, the Sm-ion's valence fluctuations\cite{Mizumaki} are not taken into account. We can, however, conclude that the magnetically robust HF compound SmOs$_4$Sb$_{12}$ at least has the magnetic field-independent relaxation of local charge degree of freedom which cause ultrasonic dispersions and will be a key to understanding the novel many-body effect in the present system.

This work was supported by Grant-in-Aid for Scientific Research on Innovative Areas `Heavy Electrons' (No. 21102501) and for Specially Promoted Research (No. 18002008) `Strongly correlated quantum phases associated with charge fluctuations' of the MEXT, Japan. Research at UCSD was supported by the U.S. Department of Energy under Grant No. DE-FG02-04ER46105 and the National Science Foundation under Grant No. DMR 0802478 for single crystal growth and characterization.  One of the authors (T.Y.) was supported by Hokkaido Univ. Leader Development System in the Basic Interdisciplinary Research Areas and (P.-C.H.) was supported by Research Corporation under CCSA Grant No. 7669.

\end{document}